\documentclass[aps,12pt,pra,showpacs,superscriptaddress,preprint,a4paper]{revtex4}

\usepackage{amsmath}
\usepackage{graphicx}
\usepackage{subfig}
\usepackage{array}
\usepackage{amsfonts}
\usepackage{epsf}
\usepackage{epsfig}
\usepackage{amssymb}
\usepackage{bbm}
\usepackage{delarray}

\catcode`ð=\active
\defð{\u{g}}
\catcode`Ð=\active
\defÐ{\u{G}}
\catcode`Ý=\active
\defÝ{\. I}
\catcode`ö=\active
\defö{\"{o}}
\catcode`Ö=\active
\defÖ{\"O}
\catcode`ü=\active
\defü{\"{u}}
\catcode`Ü=\active
\defÜ{\"{U}}
\catcode`Þ=\active
\defÞ{\c{S}}
\catcode`þ=\active
\defþ{\c{s}}
\catcode`ý=\active
\defý{{\i}}
\catcode`ç=\active
\defç{\d{c}}
\catcode`Ç=\active
\defÇ{\d{C}}

\begin{document}

\title{Exact Analytical Solution of the $N$-dimensional Radial Schrödinger Equation
with Pseudoharmonic Potential via Laplace Transform Approach}
\author{\small Tapas Das}
\email[E-mail: ]{tapasd20@gmail.com}\affiliation{Kodalia Prasanna Banga High
 School (H.S), South 24 Parganas, 700146, India}
\author{\small Altuð Arda}
\email[E-mail: ]{arda@hacettepe.edu.tr}\affiliation{Department of
Physics Education, Hacettepe University, 06800, Ankara,Turkey}
\begin{abstract}
The second order $N$-dimensional Schrödinger equation with
pseudoharmonic potential is reduced to a first order differential
equation by using the Laplace transform approach and exact bound
state solutions are obtained using convolution theorem. Some special cases are verified and variation of energy eigenvalues $E_n$ as a function of dimension $N$ are furnished. To give an extra depth of this letter, present approach is also briefly investigated for generalized Morse potential as an example.
\end{abstract}


\maketitle

\newpage

\section{I\lowercase{ntroduction}}
Schrödinger equation has long been recognized as an essential tool
for the study of atoms, nuclei, molecules and their spectral
behaviors. Much effort has been spent on to find the exact bound state solution of this non-relativistic equation for various
potentials describing the nature of bonding or the nature
of vibration of quantum systems. A large number of research
workers all around the world continue to study the ever fascinating Schrödinger equation,
which has wide application over vast areas of theoretical physics.
The Schrödinger equation is traditionally solved by
operator algebraic method [1], power series method [2-3] or path
integral method [4].

There are various other alternative methods in the literature to solve Schrödinger equation such as Fourier transform method [5-7], Nikiforov-Uvarov method [8], asymptotic iteration method [9], and SUSYQM [10]. The Laplace transformation method is also an alternative method in the list and it has a long history.
The LTA was first used by Schrödinger to derive the radial eigenfunctions of the hydrogen atom [11].
Later Englefield used LTA to solve the Coulomb, oscillator, exponential and Yamaguchi potentials [12].
Using the same methodology, the Schrödinger equation has also been solved for various other
potentials, such that pesudoharmonic [13], Dirac delta [14], Morse-type [15-16] and
harmonic oscillator [17] specially on lower dimensions.

Recently, $N$-dimensional Schrödinger equations have received focal attention in literature.
The hydrogen atom in five dimensions and isotropic oscillator in eight dimensions have been discussed by Davtyan
and co-workers [18].
Chatterjee has reviewed several methods commonly adopted for the study of $N$-dimensional Schrödinger equations
in the large $N$ limit [19], where a relevant $1/N$ expansion can be used.
Later Yanez and co-workers have investigated the position and momentum information entropies of $N$-dimensional
system [20]. The quantization of angular momentum in $N$-dimensions has been described by Al-Jaber [21].
Other recent studies of Schrödinger equation in higher dimension include
isotropic harmonic oscillator plus inverse quadratic potential [22],
$N$-dimensional radial Schrödinger equation with the Coulomb
potential [23]. Some recent works on $N$-dimensional Schrödinger equation can be found in the reference list [24-31].\\
These higher dimension studies facilitates a general treatment of the problem in such a manner that one can obtain the required results in lower dimensions just dialing appropriate $N$. The pseudoharmonic potential is expressed in the form [32]
\begin{eqnarray}
V(r)=D_e\bigg(\frac{r}{r_e}-\frac{r_e}{r}\bigg)^2\,,
\end{eqnarray}
where $D_e=\frac{1}{8}K_e r_e^{2}$ is the dissociation energy with the force constant $K_e$, and $r_e$ is the equilibrium constant. The pseudoharmonic potential is generally used to describe the roto-vibrational states of diatomic molecules and nuclear rotations and vibrations. Moreover, the pseudoharmonic potential and some kinds of it for $N$-dimensional Schrödinger equation help to test the powerfulness of different analytical methods for solving differential equations. To give an example, the dynamical algebra of the Schrödinger equation in $N$-dimension has been studied by using pseudoharmonic potential [33]. Taseli and co-workers have tested the accuracy of expanding of the eigenfunction in a Fourier-Bessel basis [34] and a Laguerre-basis [35] with the help of different type of polynomial potentials in $N$-dimension.\\
Motivated by these type of works, in this present article we discuss the exact solutions of the $N$-dimensional radial Schrödinger equation with pseudoharmonic potential using the Laplace transform approach.
To make this paper self contained we briefly outline Laplace transform method and convolution theorem in the next section. Section 3 is for the bound state spectrum of the potential system. Section 4 is devoted for the results and discussion where we derive some well known results for special cases of the potential. The application of the present method is shown in section 5 where we briefly show how generalized Morse potential could be solved. Finally the conclusion of the present work is placed in section 6.  

\section{L\lowercase{aplace} t\lowercase{ransform} m\lowercase{ethod and} c\lowercase{onvolution} t\lowercase{heorem}}
The Laplace transform $F(s)$ or $\mathcal{L}$ of a function $\chi(y)$ is defined by [36, 37]
\begin{eqnarray}
F(s)=\mathcal{L}\left\{{\chi(y)}\right\}=\int_{0}^{\infty}e^{-{sy}}{\chi(y)}dy\,.
\end{eqnarray}
If there is some constant $\sigma\in \Re$ such that ${\left|e^{-{\sigma}{y}}{\chi(y)}\right|\leq \bf M}$
for sufficiently large $y$, the integral in Eq.(2) will exist for  Re $s>\sigma$ . The Laplace transform may fail to exist because of a sufficiently strong singularity in the function $\chi(y)$ as $y\rightarrow 0$ . In particular
\begin{eqnarray}
\mathcal{L}\left[\frac{y^{\alpha}}{\Gamma(\alpha+1)}\right]=\frac{1}{s^{\alpha+1}}\,,{\alpha}>-1\,.
\end{eqnarray}
The Laplace transform has the derivative properties
\begin{eqnarray}
\mathcal{L}\left\{\chi^{(n)}(y)\right\}=s^n\mathcal{L}\left\{\chi(y)\right\}-\sum_{j=0}^{n-1}s^{n-1-j}{\chi^{(j)}(0)}\,,
\end{eqnarray}
\begin{eqnarray}
\mathcal{L}\left\{y^{n}\chi(y)\right\}=(-1)^{n}F^{(n)}(s)\,,
\end{eqnarray}
where the superscript $(n)$ denotes the $n$-th derivative with respect to $y$ for $\chi^{(n)}{(y)}$, and with respect to $s$
for $F^{(n)}{(s)}$.\\
The inverse transform is defined as $\mathcal{L}^{-1}\left\{F(s)\right\}=\chi(y)$. One of the most important properties of the Laplace transform is that given by the convolution theorem [38]. This theorem is a powerful tool to find the inverse Laplace transform. According to this theorem if we have two transformed function $g(s)=\mathcal{L}\left\{G(y)\right\}$ and
$h(s)=\mathcal{L}\left\{H(y)\right\}$, then the product of these two is the Laplace transform of the convolution $(G*H)(y)$, where
\begin{eqnarray}
(G*H)(y)=\int_{0}^y G(y-\tau)H(\tau)d\tau\,.
\end{eqnarray}
So the convolution theorem yields
\begin{eqnarray}
\mathcal{L}(G*H)(y)=g(s)h(s)\,.
\end{eqnarray}
Hence
\begin{eqnarray}
\mathcal{L}^{-1}\left\{g(s)h(s)\right\}=\int_{0}^y G(y-\tau)H(\tau)d\tau\,.
\end{eqnarray}
If we substitute $w=y-\tau$, then we find the important consequence $G*H=H*G$.

\section{B\lowercase{ound} s\lowercase{tate} s\lowercase{pectrum}}
The $N$-dimensional time-independent Schr\"{o}dinger equation for a particle of mass $M$ with orbital angular momentum quantum number
$\ell$ is given by [39]
\begin{eqnarray}
\Bigg[\nabla_N^{2}+\frac{2M}{\hbar^2}\Big(E-V(r)\Big)\Bigg]\psi_{n\ell m}(r,\Omega_N)=0\,,
\end{eqnarray}
where $E$ and  $V(r)$ denote the energy eigenvalues and potential. $\Omega_N$ within the argument of $n$-th state eigenfunctions $\psi_{n\ell m}$ denotes angular variables
$\theta_1,\theta_2,\theta_3,.....,\theta_{N-2},\varphi$. The Laplacian operator in hyperspherical coordinates is written as
\begin{eqnarray}
\nabla_N^{2}=\frac{1}{r^{N-1}}\frac{\partial}{\partial r}\left(r^{N-1}\frac{\partial}{\partial r}\right)-\frac{\Lambda_{N-1}^{2}}{r^2}\,,
\end{eqnarray}
where
\begin{eqnarray}
\Lambda_{N-1}^{2}=-\Bigg[\sum_{k=1}^{N-2}\frac{1}{sin^2\theta_{k+1}sin^2\theta_{k+2}.....sin^2\theta_{N-2} sin^2\varphi}\times\left(\frac{1}{sin^{k-1}\theta_k}\frac{\partial}{\partial \theta_k}sin^{k-1}\theta_k\frac{\partial}{\partial \theta_k}\right)\nonumber\\+\frac{1}{sin^{N-2}\varphi}\frac{\partial}{\partial\varphi}sin^{N-2}\varphi\frac{\partial}{\partial\varphi}\Bigg]\,.
\end{eqnarray}
$\Lambda_{N-1}^{2}$ is known as the hyperangular momentum operator.\\
We chose the bound state eigenfunctions $\psi_{n\ell m}(r,\Omega_N)$ that are vanishing for $r\rightarrow0$ and $r\rightarrow\infty$. Applying the separation variable method by means of the solution
$\psi_{n\ell m}(r,\Omega_N)=R_{n\ell}(r)Y_\ell^{m}(\Omega_N)$,  Eq. (9) provides two separated equations
\begin{eqnarray}
\Lambda_{N-1}^{2}Y_\ell^{m}(\Omega_N)=\ell(\ell+N-2)Y_\ell^{m}(\Omega_N)\,,
\end{eqnarray}
where $Y_\ell^{m}(\Omega_N)$ is known as the hyperspherical harmonics and the
hyperradial or in short the ``radial'' equation
\begin{eqnarray}
\left[\frac{d^2}{dr^2}+\frac{N-1}{r}\frac{d}{dr}-\frac{\ell(\ell+N-2)}{r^2}-\frac{2M}{\hbar^2}[V(r)-E]\right]R_{n\ell}(r)=0\,,
\end{eqnarray}
where $\ell(\ell+N-2)|_{N>1}$ is the separation constant [40, 41] with $\ell=0, 1,
2, \ldots$\\
In spite of taking Eq.(1) we take the more general form of pseudoharmonic potential [42]
\begin{eqnarray}
V(r)=a_{1}r^2+\frac{a_2}{r^2}+a_{3}\,,
\end{eqnarray}
where $a_1, a_2, a_3$ are three parameters can take any real value. If we set $a_1=\frac{D_e}{r_e^{2}}$, $a_2=D_er_e^{2}$ and $a_3=-2D_e$, Eq. (14) converts into the special case what we have given in Eq.(1).
Taking this into Eq. (13) and using the following abbreviations
\begin{eqnarray}
\nu(\nu+1)=\ell(\ell+N-2)+\frac{2M}{\hbar^2}a_{2}\,\,;\mu^2=\frac{2M}{\hbar^2}a_{1}\,\,;\epsilon^2=\frac{2M}{\hbar^2}(E-a_{3})\,,
\end{eqnarray}
we obtain
\begin{eqnarray}
\left[\frac{d^2}{dr^2}+\frac{N-1}{r}\frac{d}{dr}-\frac{\nu(\nu+1)}{r^2}-\mu^2r^2+\epsilon^2\right]R_{n\ell}(r)=0\,.
\end{eqnarray}
We are looking for the bound state solutions for $R_{n\ell}(r)$ with the following properties $R_{n\ell}(r\rightarrow0)=r^{-k}$ and $R_{n\ell}(r\rightarrow\infty)=0$. Let us assume a solution of type $R_{n\ell}(r)=r^{-k}f(r)$ with $k>0$. Here the term
$r^{-k}$ ensures the fact $R_{n\ell}(r\rightarrow\infty)=0$ and $f(r)$ is expected to behave like $f(r\rightarrow0)=0$.\\
Now changing the variable as $y=r^2$ and taking $f(r)\equiv \chi(y)$ from Eq. (16)
we obtain
\begin{eqnarray}
4y\frac{d^2\chi}{dy^2}+2(N-2k)\frac{d\chi}{dy}+\left[\frac{k(k+1)-k(N-1)-\nu(\nu+1)}{y}-\mu^2y+\epsilon^2\right]\chi=0\,.
\end{eqnarray}
In order to get an exact solution of the above differential equation we remove the singular term by imposing the condition
\begin{eqnarray}
k(k+1)-k(N-1)-\nu(\nu+1)=0\,.
\end{eqnarray}
So we have
\begin{eqnarray}
y\frac{d^2\chi}{dy^2}-\left(k_{\ell N}-\frac{N}{2}\right)\frac{d\chi}{dy}-\frac{1}{4}\,(\mu^2y-\epsilon^2)\chi=0\,,
\end{eqnarray}
where $k_{\ell N}$ is taken as the positive solution of $k$, which can be easily found by little algebra from Eq. (18).
It is worth to mention here that, the condition given by Eq. (18) is necessary to get an analytical solution because otherwise only approximate or numerical solution is possible.
Introducing the Laplace transform $F(s)=\mathcal{L}\{\chi(y)\}$ with the boundary condition $f(0)\equiv\chi(0)=0$ and using Eq.(4-5), Eq.   (19) can read
\begin{eqnarray}
\left(s^2-\frac{\mu^2}{4}\right)\frac{dF}{ds}+(\eta_{\ell N}s-\frac{\epsilon^2}{4})F=0\,,
\end{eqnarray}
where $\eta_{\ell N}=k_{\ell N}-\frac{N}{2}+2$.
This parameter can have integer or non integer values and there will be an integer or non integer term(s) in energy eigenvalue according to values of $\eta_{\ell N}$ which can be seen below.\\
The solution of the last equation can be written easily as
\begin{eqnarray}
F(s)=C\left(s+\frac{\mu}{2}\right)^{-\frac{\eta_{\ell N}}{2}-\frac{\epsilon^2}{4\mu}}\left(s-\frac{\mu}{2}\right)^{-\frac{\eta_{\ell N}}{2}+\frac{\epsilon^2}{4\mu}}\,,
\end{eqnarray}
where $C$ is a constant. The term $\mu=\frac{\sqrt{2Ma_1}}{\hbar}$ is a positive real number as we restrict ourselves for the choice $a_1>0$. Now, since $s$ is positive and $\mu>0$, then second factor of Eq.(21) could become negative if $\frac{\mu}{2}>s>0$ and thus its power must be a positive integer to get singled valued eigenfunctions. This will also exclude the possibility of getting singularity in the transformation. So we have
\begin{eqnarray}
\frac{\epsilon^2}{4\mu}-\frac{1}{2}\eta_{\ell N}=n\,, n=0, 1,
2, \ldots
\end{eqnarray}
Using Eq. (22), we have from Eq. (21)
\begin{eqnarray}
F(s)=C(s+\frac{\mu}{2})^{-a}(s-\frac{\mu}{2})^{-b}=Cg(s)h(s)\,,
\end{eqnarray}
where $a=\eta_{\ell N}+n$ and $b=-n$. In order to find $\chi(y)=\mathcal{L}^{-1}\left\{F(s)\right\}$, we find [43]
\begin{eqnarray}
\mathcal{L}^{-1}\left\{(s+\frac{\mu}{2})^{-a}\right\}=G(y)=\frac{y^{a-1}{e^{-\frac{\mu}{2}y}}}{\Gamma(a)}\,,\nonumber\\
\mathcal{L}^{-1}\left\{(s-\frac{\mu}{2})^{-b}\right\}=H(y)=\frac{y^{b-1}{e^{\frac{\mu}{2}y}}}{\Gamma(b)}\,.
\end{eqnarray}
Therefore using Eq. (8) and Eq. (24), we have
\begin{eqnarray}
\chi(y)=\mathcal{L}^{-1}\left\{F(s)\right\}=C(G*H)(y)=C\int_{0}^y G(y-\tau)H(\tau)d\tau\nonumber\\
=\frac{Ce^{-\frac{\mu}{2} y}}{\Gamma(a)\Gamma(b)}\int_{0}^y (y-\tau)^{a-1}\tau^{b-1} e^{\mu\tau}d\tau\,.
\end{eqnarray}
The integration can be found in [44], which gives
\begin{eqnarray}
\int_{0}^y (y-\tau)^{a-1}\tau^{b-1} e^{\mu\tau}d\tau=B(a,b)y^{a+b-1}\,_{1}F_{1}(b, a+b, \mu y)\,,
\end{eqnarray}
where $\,_{1}F_{1}$ is the confluent hypergeometric functions. Now using the Beta function \\
$B(a,b)=\frac{\Gamma (a)\Gamma(b)}{\Gamma(a+b)}$, $\chi(y)$ can be written
\begin{eqnarray}
\chi(y)=\frac{C}{\Gamma(a+b)}e^{-\frac{\mu}{2} y}y^{a+b-1}\,_{1}F_{1}(b, a+b, \mu y)\,.
\end{eqnarray}
So we have the radial eigenfunctions
\begin{eqnarray}
R_{n\ell N}(r)=r^{-k_{{\ell N}}}f(r)=C_{n\ell N}e^{-\frac{\mu}{2}r^2}r^{(\eta_{\ell N}-\frac{N}{2})}\,_{1}F_{1}(-n,\eta_{\ell N},\mu r^2)\,,
\end{eqnarray}
where $C_{n\ell N}$ is the normalization constant. It should be noted that because of the ``quantization condition'' given by
Eq. (22), the radial eigenfunctions are possible to write in a polynomial form of degree $n$, as $\,_{1}F_{1}(b, a+b, \mu y)$ converges for all finite $\mu y$ with $b=-n$ and $(a+b)$ is not a negative integer or zero. \\
Now the normalization constant $C_{n\ell N}$ can be evaluated from the condition [45]
\begin{eqnarray}
\int_{0}^{\infty}[R_{n\ell N}(r)]^2r^{N-1}dr=1\,.
\end{eqnarray}
To evaluate the integration the formula $\,_{1}F_{1}(-q,p+1,z)=\frac{q!p!}{(q+p)!}L^{p}_{q}(z)$ is useful here, where
$L^{p}_{q}(u)$ are the Laguerre polynomials. It should be remember that the formula is applicable only if $q$ is positive integer. Hence\\
Identifying $q=n,\,\,p=\eta_{\ell N}-1$, and $z=\mu r^2$ we have
\begin{eqnarray}
\,_{1}F_{1}(-n,\eta_{\ell N},\mu r^2)=\frac{n!(\eta_{\ell N}-1)!}{(\eta_{\ell N}+n-1)!}L_{n}^{(\eta_{\ell N}-1)}(\mu r^2)\,.
\end{eqnarray}
So using the following formula for the Laguerre polynomials
\begin{eqnarray}
\int_{0}^{\infty}x^{A}e^{-x}[L_{B}^{A}(x)]^2dx=\frac{\Gamma(A+B+1)}{B!}\,,
\end{eqnarray}
we write the normalization constant
\begin{eqnarray}
C_{n\ell N}=\sqrt{2\,}\mu^{\frac{1}{2}\eta_{\ell N}}\sqrt{\frac{n!}{\Gamma(\eta_{\ell N}+n)}\,}\,\frac{(\eta_{\ell N}+n-1)!}{n!(\eta_{\ell N}-1)!}\,.
\end{eqnarray}
Finally, the energy eigenvalues are obtained from Eq. (22) along with Eq. (18) and Eq. (15)
\begin{eqnarray}
E_{n\ell N}=\frac{\hbar^2}{2M}\,\epsilon^2+a_3=a_3+\sqrt{\frac{8\hbar^2a_{1}}{M}\,}\left[n+\frac{1}{2}
+\frac{1}{4}\sqrt{(N+2\ell-2)^2+\frac{8Ma_{2}}{\hbar^2}\,}\right]\,,
\end{eqnarray}
and we write the corresponding normalized eigenfunctions as
\begin{eqnarray*}
R_{n\ell N}(r)=\sqrt{2\,}\mu^{\frac{1}{2}\eta_{\ell N}}\sqrt{\frac{n!}{\Gamma(\eta_{\ell N}+n)}\,}\,\frac{(\eta_{\ell N}+n-1)!}{n!(\eta_{\ell N}-1)!}e^{-\frac{\mu}{2}r^2}r^{(\eta_{\ell N}-\frac{N}{2})}\,_{1}F_{1}(-n,\eta_{\ell N},\mu r^2)\,,
\end{eqnarray*}
or
\begin{eqnarray}
R_{n\ell N}(r)=\sqrt{2\,}\mu^{\frac{1}{2}\eta_{\ell N}}\sqrt{\frac{n!}{\Gamma(\eta_{\ell N}+n)}\,}
e^{-\frac{\mu r^2
}{2}}r^{(\eta_{\ell N}-\frac{N}{2})}\,L_{n}^{(\eta_{\ell N}-1)}(\mu r^2)\,.
\end{eqnarray}
Finally, the complete orthonormalized eigenfunctions of the
$N$-dimensional Schrödinger equation with pseudoharmonic potential
can be given by
\begin{eqnarray}
\psi(r, \theta_{1}, \theta_{2}, \ldots, \theta_{N-2},
\phi)=\sum_{n, \ell,
m}C_{n\ell N}R_{n\ell N}(r)Y_{\ell}^{m}(\theta_{1}, \theta_{2},
\ldots, \theta_{N-2}, \phi)\,.
\end{eqnarray}

\section{R\lowercase{esults} \lowercase{and} \lowercase{discussion}}
 In this section we have shown that, the results obtained in section 3 are very useful in deriving the special cases of several potentials for lower as well as for higher dimensional wave equation.
\subsection{\textit{Isotropic harmonic oscillator}}

\begin{enumerate}
\item \textit{Three dimensions (N=3)}. \\
For this case $a_1=\frac{1}{2}M\omega^2$ and $a_2=a_3=0$ which gives from Eq. (33)
\begin{eqnarray}
E_{n\ell 3}=\hbar\omega\left(2n+l+\frac{3}{2}\right)\,,
\end{eqnarray}
where $\omega$ is the circular frequency of the particle.
From Eq. (15) and Eq. (18) we obtain $k_{\ell 3}=\ell+1$. This makes $\eta_{\ell 3}=\ell+\frac{3}{2}$ and we get
radial eigenfunctions from Eq. (34). The result agrees with
those obtained in Ref. [22].

\item \textit{Arbitrary N dimensions.} \\
Here $N$ is an arbitrary constant and as before $a_1=\frac{1}{2}M\omega^2, a_2=a_3=0$. We have the energy eigenvalues from
Eq. (33)
\begin{eqnarray}
E_{n\ell N}=\hbar\omega\left(2n+\ell+\frac{N}{2}\right)\,.
\end{eqnarray}
Solving Eq. (18) with the help of Eq. (15) we have $k_{\ell N}=\ell+N-2$ and one can
easily obtain the normalization constant from Eq. (32)
\begin{eqnarray}
C_{n\ell N}=\left[\frac{2(\frac{M\omega}{\hbar})^{(\ell+\frac{N}{2})}n!}{\Gamma(\ell+\frac{N}{2}+n)}\right]^{1/2}\,
\frac{(n+\ell+\frac{N-2}{2})!}{n!(\ell+\frac{N-2}{2})!}\,,
\end{eqnarray}
where $\eta_{\ell N}=\ell+\frac{N}{2}$.
The radial eigenfunctions are given in Eq. (34) with the above
normalization constant.
The results obtained here agree with those
found in some earlier works [22, 46, 47].
\end{enumerate}

\subsection{\textit{Isotropic harmonic oscillator plus inverse quadratic potential}}
\begin{enumerate}
\item \textit{Two dimensions (N=2).} \\
Here we have $a_1=\frac{1}{2}\,M\omega^2, a_2 \neq 0$ and $a_3=0$
where $\omega$ is the circular frequency of the particle. So from
Eq. (32) we obtain
\begin{eqnarray}
C_{n\ell 2}=\left[\frac{2(\frac{M\omega}{\hbar})^{k_{\ell 2}+1}n!}{\Gamma(k_{\ell 2}+n+1)}\right]^{1/2}\,\frac{(k_{\ell 2}+n)!}{n!k_{\ell 2}!}\,,
\end{eqnarray}
where $\eta_{\ell 2}=k_{\ell 2}+1$. $k_{\ell 2}$ can be obtained from Eq. (15) and Eq. (18) as
$k_{\ell 2}=\sqrt{\nu(\nu+1)\,}=\sqrt{\ell^2+\frac{2M}{\hbar^2}\,a_{2}\,}$.
Hence Eq. (33) gives the energy eigenvalues of the system
\begin{eqnarray}
E_{n\ell 2}=\hbar\omega(2n+k_{\ell 2}+1)\,.
\end{eqnarray}
This result have already been obtained in Refs. [22, 48].
\item \textit{Three dimensions (N=3).} \\
Here  $a_1=\frac{1}{2}\,M\omega^2$, $a_2 \neq 0$ and $a_3=0$ which
gives the energy eigenvalues as
\begin{eqnarray}
E_{n\ell 3}=\frac{\hbar\omega}{2}\left[4n+2+\sqrt{(2\ell+1)^2+\frac{8Ma_{2}}{\hbar^2}\,}\right]\,.
\end{eqnarray}
Solving Eq. (18) we get $k_{\ell 3}=\nu+1$ and hence $\eta_{\ell 3}=\nu+\frac{3}{2}$. The radial eigenfunction
can hence be obtained from Eq. (34) which also corresponds to the result
obtained in Ref. [22].
\end{enumerate}

\subsection{\textit{ $3$-dimensional Schrödinger equation with pseudoharmonic potential}}
Here Eq. (18) gives $k_{\ell 3}=\nu+1 $ and this makes $\eta_{\ell 3}=\nu+\frac{3}{2}$. So Eq. (32) provides
\begin{eqnarray}
C_{n\ell 3}=\mu^{\frac{1}{2}\left(\nu+\frac{3}{2}\right)}\sqrt{\frac{2\Gamma(\nu+n+\frac{3}{2})}{n!}\,}\left[\Gamma\left(\nu+\frac{3}{2}\right)\right]^{-1}\,,
\end{eqnarray}
and hence the normalized eigenfunctions become
\begin{eqnarray}
R_{n\ell 3}(r)=\mu^{\frac{1}{2}\left(\nu+\frac{3}{2}\right)}\sqrt{\frac{2\Gamma(\nu+n+\frac{3}{2})}{n!}\,}\left[\Gamma\left(\nu+\frac{3}{2}\right)\right]^{-1}\nonumber\\ \times e^{-\frac{\mu}{2}\,r^2}r^{\nu+1}\,_{1}F_{1}(-n,\nu+\frac{3}{2},\mu r^2)\,,
\end{eqnarray}
with the energy eigenvalues
\begin{eqnarray}
E_{n\ell 3}=a_3+\frac{\hbar^2}{2M}\,\epsilon^2=a_3+\sqrt{\frac{8\hbar^2a_{1}}{M}\,}\left[n+\frac{1}{2}
+\frac{1}{4}\sqrt{(2\ell+1)^2+\frac{8Ma_{2}}{\hbar^2}\,}\right]\,.
\end{eqnarray}
This result corresponds exactly to the ones given in Ref. [13].

\section{ S\lowercase{hort} \lowercase{review of generalized} M\lowercase{orse potential}: A\lowercase{n example}}
The generalized Morse potential [49] in terms of four parameters $V_1 ,V_2, V_3$ and $\lambda(>0) $ is
\begin{eqnarray}
V(r)=V_1e^{-\lambda(r-r_e)}+V_2e^{-2\lambda(r-r_e)}+V_3\,,
\end{eqnarray}
where $\lambda$ describes the characteristic range of the potential, $r_e$ is the equilibrium molecular separation and $V_1,V_2,V_3$ are related to the potential depth. In this short section we will only show Laplace transformable differential equation like Eq.(19) can also be achieved if the exponential and singular term $\frac{1}{r}$ and $\frac{1}{r^2}$ are properly handled. Looking back to the solution given by Eq.(34) we predetermine the solution for Eq.(9) for the potential given by Eq.(45) as
\begin{eqnarray}
\psi_{n\ell m}(r,\Omega_N)=r^{-\frac{N-1}{2}}R_{n\ell}(r)Y_\ell^{m}(\Omega_N)\,.
\end{eqnarray}
This substitution facilitates following more easier differential equation (without the $\frac{1}{r}$ term) similar to Eq.(13) of section 3.
\begin{eqnarray}
\Bigg[\frac{d^2}{dr^2}+\frac{2M}{\hbar^2}\Big(E-V(r)\Big)-\frac{A_N^{\ell}}{r^2}\Bigg]R_{n\ell}(r)=0\,,
\end{eqnarray}
where $A_N^{\ell}=\frac{(N+2\ell-1)(N+2\ell-3)}{4}$.\\
let us introduce a new variable $x=\frac{r-r_e}{r_e}$. Hence inserting Eq.(45) into Eq.(47) we have
\begin{eqnarray}
\Bigg[\frac{d^2}{dx^2}+\frac{2Mr_e^2}{\hbar^2}\left\{E-V_1e^{-\alpha x}-V_2e^{-2\alpha x}-V_3\right\}+\frac{A_N^{\ell}}{(1+x)^2}\Bigg]R(x)=0\,,
\end{eqnarray}
where $\alpha=\lambda r_e$.\\
It is possible to expand the term $\frac{1}{(1+x)^2}$ in power series as within the molecular point of view $x<<1$. So
\begin{eqnarray}
\frac{1}{(1+x)^2}\approx 1-2x+3x^2-O(x^3)\,.
\end{eqnarray}
This expansion can also be rewritten as
\begin{eqnarray}
\frac{1}{(1+x)^2}\approx c_0+c_1e^{-\alpha x}+c_2e^{-2\alpha x}\,.
\end{eqnarray}
By comparing these two it is not hard to justify that \\
$c_0=1-\frac{3}{\alpha}+\frac{3}{\alpha^2}; c_1=\frac{4}{\alpha}-\frac{6}{\alpha^2}; c_2=-\frac{1}{\alpha}+\frac{3}{\alpha^2}$.\\
Inserting Eq.(50) into Eq.(48) and changing the variable as $y=e^{-\alpha x}$ one can easily get
\begin{eqnarray}
\bigg[y^2\frac{d^2}{dy^2}+y\frac{d}{dy}+\frac{\epsilon}{\alpha^2}+\frac{B_1}{\alpha^2}y+\frac{B_2}{\alpha^2}y^2\bigg] R(y)=0\,.
\end{eqnarray}
Now further assuming the solution of the above differential equation $R(y)=y^{-\omega}g(y)$ ,as we did in previous section  with proper requirement for bound state scenario, finally we can construct the following differential equation just like Eq.(17) for the perfect platform for Laplace transformation.
\begin{eqnarray}
y\frac{d^2g}{dy^2}+(1-2\omega)\frac{dg}{dy}+\frac{\omega^2-\frac{\epsilon}{\alpha^2}}{y}g+\Big(\frac{B_1}{\alpha^2}+\frac{B_3}{\alpha^2}y\Big)g=0\,,
\end{eqnarray}
where
\begin{subequations}
\begin{align}
\epsilon&=\frac{2Mr_e^2}{\hbar^2}(E-V_3)+c_0A_N^{\ell}\\
B_1&=c_1A_N^{\ell}-\frac{2Mr_e^2}{\hbar^2}V_1\\
B_2&=c_2A_N^{\ell}-\frac{2Mr_e^2}{\hbar^2}V_2
\end{align}
\end{subequations}
Imposing the condition $\omega^2-\frac{\epsilon}{\alpha^2}=0$ as previous and approaching the same way as we did in section 3 one can obtain the energy eigenvalues and bound state wave functions in terms of confluent hypergeometric function. We have investigated that the results are exact match with the reference [15] for $V_1=-2D, V_2=D$ and $V_3=0 $ where $D$ describes the depth of the potential.
\section{C\lowercase{onclusions}}
We have investigated some aspects of $N$-dimensional hyperradial
Schrödinger equation for pseudoharmonic potential by Laplace transformation
approach. It is found that the energy eigenfunctions and the
energy eigenvalues depend on the dimensionality of the problem. In this connection we have furnished few plot of the energy spectrum $E(N)$ as a function of $N$ for a given set of physical parameters $a_1,a_2,\ell$ and $n=0,1,2,3$ keeping $a_3=0$.The general results obtained in this article have been verified with earlier
reported results, which were obtained for certain special values of potential parameters and dimensionality.\\
The Laplace transform is a powerful, efficient and accurate alternative method of deriving energy eigenvalues and eigenfunctions of some spherically symmetric potentials that are analytically solvable. It may be hard to predict which kind of potentials are solvable analytically by Laplace transform, but in general prediction of the eigenfunctions with the form like $R_{n\ell}(r)=r^{-k}f(r)$ always open the gate of the possibility of closed form solutions for a particular potential model. This kind of substitution is called \textit{Universal Laplace transformation scheme}.In this connection one might go through the reference [50] to check out how Laplace transformation technique behaves over different potentials, specially for Schr\"{o}dinger equation in lower dimensional domain. It is also true that, the technique of Laplace transformation is useful if, inserting the potential into the Schr\"{o}dinger equation and using \textit{Universal Laplace transformation scheme} via some suitable parametric restrictions, one able to get a differential equation with variable coefficient $r^j (j\ngtr 1)$. This is not easy to achieve every time. However, for a given potential if there is no such achievement, iterative approach facilitates a better way to overcome the situation [51]. The results are sufficiently accurate for such special potentials at least for practical purpose.\\
Before concluding we want to mention here that, we have not succeed to develop the scattering state solution for the pseudoharmonic potential. If we could develop those solutions using the LTA that would have been a remarkable achievement. The main problem we have faced to handle the Eq.(21) with complex index as for scattering state situation, that means  
$E>\lim_{r\to\infty}V(r)$, $\mu$ should be replaced with $\mu=i\zeta$ where $\zeta=\sqrt{\frac{2Ma_1}{\hbar^2}}$. May be this cumbersome situation would attract the researchers to study the subject further and we are looking forward to it.        
\section*{Acknowledgements}
The author A. A. thanks Dr Andreas Fring from City University London, and the Department of Mathematics for hospitality where final version of this work has been completed. The authors thank to kind referee for valuable suggestions which have improved the manuscript greatly. Finally the author T.D wishes to dedicate this work to his wife Sonali for her love and care.

\section*{R\lowercase{eferences}}

\newpage


\end{document}